\documentclass[3p,times,twocolumn]{elsarticle}

\usepackage{ecrc}

\volume{00}

\firstpage{1}

\journalname{Nuclear Physics B Proceedings Supplement}

\runauth{E. Meggiolaro}

\jid{nuphbp}

\jnltitlelogo{Nuclear Physics B Proceedings Supplement}

\usepackage{amssymb}
\usepackage{amsfonts,amsmath}

\usepackage[figuresright]{rotating}

\newcommand{\La}{\mathcal{L}}

\newcommand{\tc}{T_c}
\newcommand{\tro}{T_{\rho_{\pi}}}
\newcommand{\tuone}{T_{U(1)}}

\newcommand{\ov}[1]{\overline{#1}}
\newcommand{\gr}[1]{\textbf{#1}}

\newcommand{\lsigma}{\mathcal{L}_{0}(U,U^{\dagger})}
\newcommand{\ropi}{\rho_{\pi}}
\newcommand{\rox}{\rho_{X}\,}

\newcommand{\lpq}{\lambda_{\pi}^{2}}

\newcommand{\bpq}{B_{\pi}^{2}}
\newcommand{\lxq}{\lambda_{X}^{2}}
\newcommand{\fxq}{F_{X}^{2}}
\newcommand{\rad}{\sqrt}

\newcommand{\Gpi}{\mathcal{G}_{\pi}}

\newcommand{\smin}{\ov{\sigma}}
\newcommand{\dmin}{\ov{\delta}}

\newcommand{\Tr}{\mathrm{Tr}}

\begin{document}

\begin{frontmatter}



\dochead{}

\title{Comments on the $U(1)$ axial symmetry and the chiral transition in QCD}


\author{Enrico Meggiolaro}
\ead{enrico.meggiolaro@df.unipi.it}

\address{Dipartimento di Fisica, Universit\`a di Pisa,
and INFN, Sezione di Pisa, Largo Pontecorvo 3, I-56127 Pisa, Italy}

\begin{abstract}
We analyze (using a chiral effective Lagrangian model) the scalar and
pseudoscalar meson mass spectrum of QCD at finite temperature, above
the chiral transition at $T_c$, looking, in particular, for signatures
of a possible breaking of the $U(1)$ axial symmetry above $T_c$.
A detailed comparison between the case with a number of light quark flavors
$N_f\geq 3$ and the (remarkably different) case $N_f=2$ is performed.
\end{abstract}


\end{frontmatter}



\section{Introduction}

The so-called {\it chiral condensate}, $\langle \bar{q}q \rangle
\equiv \sum_{l=1}^{N_f} \langle \bar{q}_l q_l \rangle$, is known to be
an order parameter for the $SU(N_f) \otimes SU(N_f)$ chiral symmetry of
the QCD Lagrangian with $N_f$ massless quarks ({\it chiral limit}),
the physically relevant cases being $N_f=2$ and $N_f=3$.
Lattice determinations of $\langle \bar{q}q \rangle$ (see,
e.g., Refs. \cite{HotQCD}) show that there is a {\it chiral phase transition}
at a temperature $\tc \sim 150 \div 170$ MeV, which is practically equal to
the {\it deconfinement} temperature $T_d$, separating the {\it confined}
(or {\it hadronic}) phase at $T<T_d$, from the {\it deconfined} phase
(also known as {\it quark-gluon plasma}) at $T>T_d$.
For $T < \tc \sim T_d$, the chiral condensate $\langle \bar{q}q \rangle$ is
nonzero and the chiral symmetry is spontaneously broken down to the vectorial
subgroup $SU(N_f)_V$, and the $N_f^2-1$ $J^P=0^-$ lightest mesons
are just the (pseudo-)Goldstone bosons associated with this breaking.
Instead, for $T > \tc \sim T_d$, the chiral condensate
$\langle \bar{q}q \rangle$ vanishes and the chiral symmetry is restored.
But this is not the whole story, since QCD with $N_f$ massless quarks
also has a $U(1)$ axial symmetry [$U(1)_A$], which
is broken by an anomaly at the quantum level \cite{Weinberg1975,tHooft1976}:
this anomaly plays a fundamental role in explaining the large mass of the
$\eta'$ meson \cite{Witten1979,Veneziano1979}.

Now, the question is: What is the role of the $U(1)$ axial symmetry for
the finite temperature phase structure of QCD?
One expects that, at least for $T \gg T_c$,  where the density
of {\it instantons} is strongly suppressed due to a Debye-type screening
\cite{GPY1981}), also the $U(1)$ axial symmetry will be ({\it effectively})
restored.
This question is surely of phenomenological relevance since the particle
mass spectrum above $T_c$ drastically depends on the presence or absence of the
$U(1)$ axial symmetry. From the theoretical point of view, this question
can be investigated by comparing (e.g., on the lattice) the behavior at
nonzero temperatures of the two-point correlation functions
$\langle O_f(x) O_f^\dagger(0) \rangle$
for the various $q\bar{q}$ meson channels (``$f$'').
For example, for $N_f=2$ \cite{DK1987,Shuryak1994}, one can study the
meson channels (traditionally called $\sigma$, $\vec\delta$, $\eta$ and
$\vec\pi$) which are listed in Table \ref{table1}, together with their
corresponding interpolating operators and their isospin ($I$) and
spin-parity ($J^P$) quantum numbers.
\begin{table}[htbp]
\begin{center}
\begin{tabular}{|l|c|c|r|}
\hline
Meson channel & Interpolating operator & $I$ & $J^{P}$ \\
\hline
$\sigma$ (or $f_0$) & $O_\sigma=\ov{q}q$ & 0 & $0^{+}$ \\
\hline
$\vec\delta$ (or $\vec{a}_0$) & $\vec{O}_\delta= \ov{q}\frac{\vec{\tau}}{2}q$
& 1 & $0^{+}$ \\
\hline
$\eta$ & $O_\eta=i\ov{q}\gamma_5 q$ & 0 & $0^{-}$ \\
\hline
$\vec\pi$ & $\vec{O}_\pi=i\ov{q}\gamma_5\frac{\vec{\tau}}{2}q$ &
1 & $0^{-}$ \\
\hline
\end{tabular}
\end{center}
\caption{$q\bar{q}$ meson channels (for $N_f=2$) and their quantum numbers.}
\label{table1}
\end{table}
Under $SU(2)_A$ and $U(1)_A$ transformations, the $q\bar{q}$ meson channels
are mixed as follows:
\begin{equation}
\begin{matrix}
\sigma & \stackrel{U(1)_A}{\longleftrightarrow} & \eta \\
SU(2)_A \updownarrow & & \updownarrow SU(2)_A \\
\vec{\pi} & \stackrel{U(1)_A}{\longleftrightarrow} & \vec{\delta}
\end{matrix}
\end{equation}
The restoration of the $SU(2)$ chiral symmetry implies that the $\sigma$ and
$\vec\pi$ channels become degenerate, with identical correlators and,
therefore, with identical ({\it screening}) masses, $M_\sigma = M_\pi$.
The same happens also for the channels $\eta$ and $\vec\delta$.
Instead, an {\it effective restoration} of the $U(1)$ axial symmetry should
imply that $\sigma$ becomes degenerate with $\eta$, and $\vec\pi$ becomes
degenerate with $\vec\delta$.
(Clearly, if both chiral symmetries are restored, then all $\sigma$, $\vec\pi$,
$\eta$, and $\vec\delta$ correlators should become the same.)

In Ref. \cite{MM2013} the scalar and pseudoscalar meson mass spectrum,
above the chiral transition at $\tc$, has been analyzed using, instead,
a chiral effective Lagrangian model (which was originally proposed in
Refs. \cite{EM1994a,EM1994b,EM1994c} and elaborated on in Refs.
\cite{MM2003,EM2004,EM2011}), which, in addition to the usual
chiral condensate $\langle \bar{q} q \rangle$, also includes a (possible)
{\it genuine} $U(1)_A$-breaking condensate that (possibly) survives across
the chiral transition at $T_c$, staying different from zero at $T > T_c$.
The motivations for considering this Lagrangian (and a critical comparison
with other effective Lagrangian models existing in the literature) are
recalled in Sec. 2.
The results for the mesonic mass spectrum for $T > T_c$ are summarized in
Sec. 3, for the case $N_f\geq 3$, and in Sec. 4, for the case $N_f=2$.
Finally, in Sec. 5, we shall make some comments on (i) the remarkable
difference between the case $N_f\geq 3$ and the case $N_f=2$, and (ii) the
comparison between our results and the available lattice results for $N_f=2$
(or $N_f=2+1$).

\section{Chiral effective Lagrangians}

Chiral symmetry restoration at nonzero temperature is often studied in the
framework of the following effective Lagrangian
\cite{PW1984,tHooft1986,LRS2000,RRR2003,Vicari-et-al.},
written in terms of the (quark-bilinear) mesonic effective field
$U_{ij} \sim \ov{q}_{jR} q_{iL} =
\ov{q}_{j}\left(\frac{1+\gamma_5}{2}\right)q_{i}$,\footnote{We use the
following notation for the left-handed and right-handed quark fields:
$q_{L,R} \equiv \frac{1}{2} (1 \pm \gamma_5) q$,
with $\gamma_5 \equiv -i\gamma^0\gamma^1\gamma^2\gamma^3$.}
\begin{eqnarray}\label{lageffstandard}
\lefteqn{
\mathcal{L}_1(U,U^\dagger) = \mathcal{L}_0(U,U^{\dagger})+
\frac{B_{m}}{2\sqrt2}\Tr[MU+M^{\dagger}U^{\dagger}] } \nonumber \\
& & ~~~~~~~~~~~~+ \mathcal{L}_{I}(U,U^\dagger),
\end{eqnarray}
where $M={\rm diag}(m_1,\dots,m_{N_f})$ is the quark mass matrix and
$\mathcal{L}_0(U,U^{\dagger})$ is a term describing a kind of linear sigma
model,
\begin{equation}\label{sigmamodel}
\begin{split}
\mathcal{L}_0(U,U^{\dagger}) =&~
\frac{1}{2}\Tr[\partial_{\mu}U\partial^{\mu}U^{\dagger}]- V_0(U,U^\dagger), \\
V_0(U,U^\dagger) =&~
\frac{1}{4}\lambda_{\pi}^{2}\Tr[(UU^{\dagger}-\rho_{\pi}\mathbf{I})^{2}]
+\frac{1}{4}\lambda_{\pi}^{\prime 2}[\Tr(UU^{\dagger})]^{2} ,
\end{split}
\end{equation}
while $\mathcal{L}_{I}(U,U^\dagger)$ is an interaction term of the form:
\begin{equation}\label{thooftterm}
\mathcal{L}_{I}(U,U^\dagger)=c_{I}[\det U+\det U^{\dagger}].
\end{equation}
Since under $U(N_f)_L\otimes U(N_f)_R$ chiral transformations the quark fields
and the mesonic effective field $U$ transform as
\begin{equation}\label{trasfU}
U(N_f)_L\otimes U(N_f)_R:~ q_{L,R}\rightarrow V_{L,R}q_{L,R} ~\Rightarrow~
U\rightarrow V_{L} U V_{R}^{\dagger} ,
\end{equation}
where $V_L$ and $V_R$ are arbitrary $N_f \times N_f$ unitary matrices,
we have that $\lsigma$ is invariant under the entire chiral group
$U(N_f)_L\otimes U(N_f)_R$, while the interaction term \eqref{thooftterm} [and
so the entire effective Lagrangian \eqref{lageffstandard} in the chiral limit
$M=0$] is invariant under $SU(N_f)_L\otimes SU(N_f)_R \otimes U(1)_V$ but
{\it not} under a $U(1)$ axial transformation:
\begin{equation}\label{U1trasf}
U(1)_A:~ q_{L,R}\rightarrow e^{\mp i\alpha}q_{L,R} ~\Rightarrow~
U \rightarrow e^{-i2\alpha} U .
\end{equation}
However, as was noticed by Witten \cite{Witten1980}, Di Vecchia, and Veneziano
\cite{DV1980}, this type of {\it anomalous} term does not correctly
reproduce the U(1) axial anomaly of the fundamental theory, i.e., of the QCD
(and, moreover, it is inconsistent with the $1/N_c$ expansion).
In fact, one should require that, under a $U(1)$ axial transformation
\eqref{U1trasf}, the effective Lagrangian, in the chiral limit $M=0$,
transforms as
\begin{equation}\label{trasfanomal}
U(1)_A:~ \mathcal{L}^{(M=0)}_{eff} \rightarrow
\mathcal{L}^{(M=0)}_{eff} + \alpha 2N_f Q ,
\end{equation}
where $Q(x)=\frac{g^{2}}{64\pi^{2}}\varepsilon^{\mu\nu\rho\sigma}
F_{\mu\nu}^{a}(x)F_{\rho\sigma}^{a}(x)$ is the {\it topological charge density}
and $\mathcal{L}_{eff}$ also contains $Q$ as an auxiliary field.
The correct effective Lagrangian, satisfying the transformation property
\eqref{trasfanomal}, was derived in Refs.
\cite{Witten1980,DV1980,RST1980,KO1980,NA1981} and is given by
\begin{eqnarray}\label{lageff}
\lefteqn{
\hspace{-0.5cm}
\mathcal{L}_2(U,U^{\dagger},Q) = \mathcal{L}_0(U,U^{\dagger})
+\frac{B_{m}}{2\sqrt2}\Tr[MU+M^{\dagger}U^{\dagger}] } \nonumber \\
& & ~+ \frac{i}{2}Q\Tr[\log U-\log U^{\dagger}]+\frac{1}{2A}Q^{2} ,
\end{eqnarray}
where $A=-i\int d^{4}x \langle TQ(x)Q(0)\rangle|_{YM}$ is the so-called
{\it topological susceptibility} in the pure Yang--Mills (YM) theory.
After integrating out the variable $Q$ in the effective Lagrangian
\eqref{lageff}, we are left with
\begin{eqnarray}\label{lageffint}
\lefteqn{
\mathcal{L}_2(U,U^{\dagger})=\mathcal{L}_{0}(U,U^{\dagger})
+\frac{B_{m}}{2\sqrt2}\Tr[MU+M^{\dagger}U^{\dagger}] } \nonumber \\
& & ~+ \frac{1}{8}A\left\{\Tr[\log U-\log U^{\dagger}]\right\}^2 ,
\end{eqnarray}
to be compared with Eqs. \eqref{lageffstandard}--\eqref{thooftterm}.

For studying the phase structure of the theory at finite temperature $T$,
all the parameters appearing in the effective Lagrangian must be considered as
functions of $T$. In particular, the parameter
$\rho_\pi$, appearing in the first term of the potential $V_0(U,U^\dagger)$
in Eq. \eqref{sigmamodel}, is responsible for the behavior of the theory
across the chiral phase transition at $T=T_c$.
Let us consider, for a moment, only the linear sigma model $\lsigma$, i.e.,
let us neglect both the anomalous symmetry-breaking term and the mass
term in Eq. \eqref{lageffint}.
If $\rho_\pi(T<T_c) > 0$, then the value $\ov{U}$ for which the potential
$V_0$ is minimum (that is, in a mean-field approach, the {\it vacuum
expectation value} of the mesonic field $U$) is different from zero and
can be chosen to be
\begin{equation}\label{UWDVminore}
\ov U\vert_{\ropi >0}=v\mathbf{I}, ~~~~
v\equiv\frac{F_\pi}{\sqrt{2}} = \sqrt{\frac{\ropi\lambda_\pi^2}
{\lambda_\pi^2+N_f\lambda_\pi^{\prime 2}}} ,
\end{equation}
which is invariant under the vectorial $U(N_f)_V$ subgroup; the chiral symmetry
is thus spontaneously broken down to $U(N_f)_V$.
Instead, if $\rho_\pi(T>T_c) < 0$, we have that
\begin{equation}\label{UWDVmaggiore}
\ov{U}\vert_{\ropi<0}=0 ,
\end{equation}
and the chiral symmetry is realized {\it \`a la} Wigner--Weyl. The critical
temperature $T_c$ for the chiral phase transition is thus, in this case,
simply the temperature at which the parameter $\rho_\pi$ vanishes:
$\rho_\pi(T_c)=0$.

However, the anomalous term in Eq. \eqref{lageffint} makes sense
only in the low-temperature phase ($T<T_c$), and it is singular for
$T>T_c$, where the vacuum expectation value of the mesonic field $U$
vanishes. On the contrary, the interaction term \eqref{thooftterm}
behaves well both in the low- and high-temperature phases.

The above-mentioned problems can be overcome by considering a {\it modified}
effective Lagrangian (which was originally proposed in
Refs. \cite{EM1994a,EM1994b,EM1994c} and elaborated on in Refs.
\cite{MM2003,EM2004,EM2011}), which, in a sense, is an ``extension''
of both $\mathcal{L}_1$ and $\mathcal{L}_2$, having
(i) the correct transformation property \eqref{trasfanomal}
under the chiral group, and (ii) an interaction term containing the
determinant of the mesonic field $U$, of the kind of that in
Eq. \eqref{thooftterm}, assuming that there is a
$U(1)_A$-breaking condensate that (possibly) survives across the chiral
transition at $T_c$, staying different from zero up to a temperature
$\tuone > T_c$. (Of course, it is also possible that $\tuone \to \infty$,
as a limit case. Another possible limit case, i.e., $\tuone = \tc$, will be
discussed in the concluding comments in Sec. 5.)
The new $U(1)$ chiral condensate has the form
$C_{U(1)} = \langle {\cal O}_{U(1)} \rangle$,
where, for a theory with $N_f$ light quark flavors, ${\cal O}_{U(1)}$ is a
$2N_f$-quark local operator that has the chiral transformation properties of
\cite{tHooft1976,KM1970,Kunihiro2009}
${\cal O}_{U(1)} \sim {\det}(\bar{q}_{sR}q_{tL}) + {\det}(\bar{q}_{sL}q_{tR})$,
where $s,t = 1, \dots ,N_f$ are flavor indices. The color indices (not
explicitly indicated) are arranged in such a way that
(i) ${\cal O}_{U(1)}$ is a color singlet, and (ii)
$C_{U(1)} = \langle {\cal O}_{U(1)} \rangle$ is a {\it genuine} $2N_f$-quark
condensate, i.e., it has no {\it disconnected} part proportional to some
power of the quark-antiquark chiral condensate $\langle \bar{q} q \rangle$;
the explicit form of the condensate for the cases $N_f=2$ and $N_f=3$ is
discussed in detail in the Appendix A of Ref. \cite{EM2011} (see also Refs.
\cite{EM1994c,DM1995}).

The modified effective Lagrangian is written in terms of the topological charge
density $Q$, the mesonic field $U_{ij} \sim \bar{q}_{jR} q_{iL}$,
and the new field variable $X \sim {\det}(\bar{q}_{sR}q_{tL})$,
associated with the $U(1)$ axial condensate \cite{EM1994a,EM1994b,EM1994c},
\begin{eqnarray}\label{lagtot}
\lefteqn{
\La (U,U^\dagger,X,X^\dagger,Q)
= \frac{1}{2}\Tr[\partial_\mu U\partial^\mu U^\dagger]
+ \frac{1}{2}\partial_\mu X\partial^\mu X^\dagger } \nonumber \\
& & \hspace{-0.1cm} -\ V(U,U^\dagger,X,X^\dagger)
+ \frac{i}{2}\omega_1 Q \Tr[\log U - \log U^\dagger] \nonumber \\
& & \hspace{-0.1cm} +\ \frac{i}{2}(1-\omega_1)Q[\log X-\log X^\dagger]
+ \frac{1}{2A}Q^2,
\end{eqnarray}
where the potential term $V(U,U^{\dagger},X,X^{\dagger})$ has the form
\begin{eqnarray}\label{V}
\lefteqn{
V(U,U^{\dagger},X,X^{\dagger}) } \nonumber \\
& & \hspace{-0.1cm} =
\frac{1}{4}\lambda_{\pi}^{2}\Tr[(UU^{\dagger}-\rho_{\pi}{\bf I})^{2}]
+\frac{1}{4}\lambda_{\pi}^{\prime 2}[\Tr(UU^{\dagger})]^{2} \nonumber \\
& & \hspace{-0.1cm} +\
\frac{1}{4}\lambda_{X}^{2}[XX^{\dagger}-\rho_{X}]^{2}
-\frac{B_{m}}{2\sqrt{2}}\Tr[MU+M^{\dagger}U^{\dagger}] \nonumber \\
& & \hspace{-0.1cm} -\
\frac{c_{1}}{2\sqrt{2}}[X^{\dagger}\det U+X\det U^{\dagger}].
\end{eqnarray}
Since under chiral $U(N_f)_L\otimes U(N_f)_R$ transformations [see Eq.
\eqref{trasfU}] the field $X$ transforms exactly as $\det U$,
\begin{equation}\label{trasfX}
U(N_f)_L\otimes U(N_f)_R:~ X \rightarrow \det V_L (\det V_R)^* X,
\end{equation}
[i.e., $X$ is invariant under $SU(N_f)_L\otimes SU(N_f)_R\otimes U(1)_V$,
while, under a $U(1)$ axial transformation \eqref{U1trasf},
$X\rightarrow e^{-i2N_f\alpha}X$],
we have that, in the chiral limit $M=0$, the effective Lagrangian
\eqref{lagtot} is invariant under $SU(N_f)_L\otimes SU(N_f)_R \otimes U(1)_V$,
while under a $U(1)$ axial transformation, it correctly transforms as
in Eq. \eqref{trasfanomal}.

After integrating out the variable $Q$ in the effective Lagrangian
\eqref{lagtot}, we are left with
\begin{eqnarray}\label{lagtotb}
\lefteqn{
\mathcal{L}(U,U^{\dagger},X,X^{\dagger})
= \frac{1}{2}\Tr[\partial_{\mu}U\partial^{\mu}U^{\dagger}]
+ \frac{1}{2}\partial_{\mu}X\partial^{\mu}X^{\dagger} } \nonumber \\
& & ~~~~~~~~~~~~~~~~~~~~- \tilde{V}(U,U^{\dagger},X,X^{\dagger}) ,
\end{eqnarray}
where
\begin{eqnarray}\label{Vtilde}
\lefteqn{
\tilde{V} = V -\frac{1}{8}A\{\omega_{1}\Tr[\log U-\log U^{\dagger}] }
\nonumber \\
& & +\ (1-\omega_{1})[\log X-\log X^{\dagger}]\}^2 .
\end{eqnarray}
As we have already said, all the parameters appearing in the effective
Lagrangian must be considered as functions of the physical temperature $T$.
In particular, the parameters $\rho_{\pi}$ and $\rho_X$ determine the
expectation values $\langle U \rangle$ and $\langle X \rangle$, and so they
are responsible for the behavior of the theory across the
$SU(N_f) \otimes SU(N_f)$ and the $U(1)$ chiral phase transitions.
We shall assume that the parameters $\rho_\pi$ and $\rho_X$, as functions
of the temperature $T$, behave as reported in Table \ref{table2};
$\tro$ is thus the temperature at which the parameter $\ropi$ vanishes, while
$\tuone>\tro$ is the temperature at which the parameter $\rox$ vanishes
(with, as we have said above, $\tuone\to\infty$, i.e., $\rox>0$ $\forall T$,
as a possible limit case).
\begin{table}[htbp]
\begin{center}
\begin{tabular}{|l|c|c|r|}
\hline
$T<\tro$ & $\tro<T<\tuone$ & $T>\tuone$ \\ 
\hline
$\ropi>0$ & $\ropi<0$ & $\ropi<0$ \\ 
\hline
$\rox>0$ & $\rox>0$ & $\rox<0$ \\ 
\hline
\end{tabular}
\end{center}
\caption{Dependence of the parameters $\ropi$, $\rox$ on the
temperature $T$.}
\label{table2}
\end{table}
We shall see in the next section that, in the case $N_f\geq 3$, one has
$\tc=\tro$ (exactly as in the case of the linear sigma model $\mathcal{L}_0$
discussed above), while, as we shall see in Sec. 4, the situation in which
$N_f=2$ is more complicated, being $\tro<\tc<\tuone$ in that case (unless
$\tro=\tc=\tuone$; this limit case will be discussed in the concluding
comments in Sec. 5).

Concerning the parameter $\omega_{1}$, in order to avoid a singular behavior of
the anomalous term in Eq. \eqref{Vtilde} above the chiral transition temperature
$T_c$, where the vacuum expectation value of the mesonic field $U$
vanishes (in the chiral limit $M=0$), we shall assume that
$\omega_{1}(T\geq T_c)=0$.

Finally, let us observe that the interaction term between the
$U$ and $X$ fields in Eq. \eqref{V}, i.e.,
\begin{equation}\label{Lint}
\mathcal{L}_{int}=\frac{c_{1}}{2\rad2}[X^{\dagger}\det U+X\det U^{\dagger}] ,
\end{equation}
is very similar to the interaction term \eqref{thooftterm}
that we have discussed above for the effective Lagrangian $\mathcal{L}_1$.
However, the term \eqref{Lint} is {\it not anomalous}, being invariant
under the chiral group $U(N_f)_L \otimes U(N_f)_R$, by virtue of Eqs.
\eqref{trasfU} and \eqref{trasfX}.
Nevertheless, if the field $X$ has a (real) {\it nonzero} vacuum expectation
value $\ov X$ [the $U(1)$ axial condensate], then we can write
\begin{equation}\label{hXSx}
X=(\ov{X}+h_X)e^{i\frac{S_X}{\ov{X}}} ~~~~
({\rm with:}~~ \ov{h}_X = \ov{S}_X = 0),
\end{equation}
and, after susbstituting this in Eq.  \eqref{Lint} and expanding in powers of
the excitations $h_X$ and $S_X$, one recovers, at the leading order, an
interaction term of the form \eqref{thooftterm}:
\begin{equation}\label{Lint-expanded}
\mathcal{L}_{int} = c_{I}[\det U+\det U^{\dagger}] + \dots ,~~~~
c_{I} \equiv \frac{c_1\ov X}{2\rad2}.
\end{equation}
In what follows (see Ref. \cite{MM2013} for more details) we shall analyze
the effects of assuming a nonzero value of the $U(1)$ axial condensate
$\ov{X}$ on the scalar and pseudoscalar meson mass spectrum {\it above}
the chiral transition temperature ($T>T_c$), both for the case $N_f\geq 3$
(Sec. 3) and for the case $N_f=2$ (Sec. 4).

\section{Mass spectrum for $T>T_c$ in the case $N_f\geq 3$}

Let us suppose to be in the range of temperatures $\tro<T<\tuone$, where,
according to Table \ref{table2},
\begin{equation}\label{Trange}
\ropi\equiv-\frac{1}{2}B_\pi^2<0,~~~~ \rox\equiv\frac{1}{2}F_X^2>0 .
\end{equation}
Since we expect that, due to the sign of the parameter $\rho_X$ in the
potential \eqref{V}, the $U(1)$ axial symmetry is broken by a nonzero
vacuum expectation value of the field $X$ (at least for $\lxq \to \infty$
we should have $\ov X^\dagger \ov X \to \frac{1}{2} F_X^2$),
we shall use for the field $U$ a simple linear parametrization,
while, for the field $X$, we shall use a nonlinear parametrization
(in the form of a {\it polar decomposition}),
\begin{equation}\label{paramUXL3}
U_{ij}=a_{ij}+ib_{ij}, \,\,\,\,X=\alpha e^{i\beta}=(\ov{\alpha}+h_X)
e^{i\left(\ov{\beta}+\frac{S_X}{\ov{\alpha}}\right)},
\end{equation}
where $\ov{X}=\ov{\alpha}e^{i\ov{\beta}}$ (with $\ov\alpha \neq 0$)
is the vacuum expectation value of $X$ and $a_{ij}$, $b_{ij}$, $h_X$, and
$S_X$ are real fields.
Inserting Eq. \eqref{paramUXL3} into the expressions \eqref{V} and
\eqref{Vtilde}, we find the expressions for the potential with and without the
anomalous term (with $\omega_1=0$),
\begin{equation}\label{Vanom}
\tilde{V} = V-\frac{1}{8}A[\log X-\log X^{\dagger}]^{2}
= V+\frac{1}{2}A\beta^2 ,
\end{equation}
\begin{equation}\label{VL3}
\nonumber
\begin{split}
V=&~\frac{1}{4}\lpq \Tr[(UU^\dagger)(UU^\dagger)]
+\frac{1}{4}\lambda_{\pi}^{\prime 2}[\Tr(UU^{\dagger})]^{2} \\
&+\frac{1}{4}\lpq\bpq(a_{ij}^2+b_{ij}^2)
+\frac{1}{4}\lxq\left(\alpha^2-\frac{1}{2}F_X^2\right)^{2} \\
&-\frac{B_m}{\rad2} m_i a_{ii}
-\frac{c_{1}}{2\rad2}[\alpha\cos\beta(\det U+\det U^\dagger) \\
&+i\alpha\sin\beta(\det U-\det U^\dagger)] +\frac{N_f}{16}\lpq B_\pi^4 .
\end{split}
\end{equation}
At the {\it minimum} of the potential we find that, at the
leading order in $M = {\rm diag}(m_1,\dots,m_{N_f})$:
\begin{equation}\label{minUalphabetaL3}
\ov{U}=\frac{2B_{m}}{\sqrt{2}\lambda_{\pi}^{2}B_{\pi}^{2}}M+\dots,~~
\ov\alpha = \frac{F_X}{\rad2} + {\cal O}(\det M),~~ \ov\beta = 0.
\end{equation}
In particular, in the chiral limit $M=0$, we find that
$\ov U = 0$ and $\ov X = \ov\alpha = \frac{F_X}{\rad2}$,
which means that, in this range of temperatures $\tro<T<\tuone$, the
$SU(N_f)_L\otimes SU(N_f)_R$ chiral symmetry is restored so that we can say
that (at least for $N_f\geq 3$) $\tc\equiv\tro$, while the $U(1)$ axial
symmetry is broken by the $U(1)$ axial condensate $\ov X$.
Concerning the mass spectrum of the effective Lagrangian, we have $2N_f^2$
degenerate scalar and pseudoscalar mesonic excitations, described by the fields
$a_{ij}$ and $b_{ij}$, {\it plus} a scalar ($0^+$) singlet field
$h_X = \alpha - \ov\alpha$ and a pseudoscalar ($0^-$) singlet field
$S_X = \ov\alpha \beta$ [see Eq. \eqref{paramUXL3}],
with squared masses given by
\begin{equation}\label{massesL3}
M^2_U = \frac{1}{2}\lpq\bpq,~~~ M^2_{h_X}=\lxq\fxq,~~~
M^2_{S_X} = \frac{A}{\ov{X}^2} = \frac{2A}{\fxq}.
\end{equation}
While the mesonic excitations described by the field $U$ are of the usual
$q \bar q$ type, the scalar singlet field $h_X$ and the pseudoscalar singlet
field $S_X$ describe instead two {\it exotic}, $2N_f$-quark excitations
of the form
$h_X ~(\alpha) \sim {\det}(\bar{q}_{sL}q_{tR}) + {\det}(\bar{q}_{sR}q_{tL})$
and $S_X \sim i[ {\det}(\bar{q}_{sL}q_{tR}) - {\det}(\bar{q}_{sR}q_{tL})]$.
In particular, the physical interpretation of the pseudoscalar singlet
excitation $S_X$ is rather obvious, and it was already discussed in Ref.
\cite{EM1994a}: it is nothing but the {\it would-be} Goldstone particle
coming from the breaking of the $U(1)$ axial symmetry. In fact, neglecting 
the anomaly, it has zero mass in the chiral limit of zero quark masses.
Yet, considering the anomaly, it acquires a {\it topological} squared
mass proportional to the topological susceptibility $A$ of the pure YM theory,
as required by the Witten--Veneziano mechanism \cite{Witten1979,Veneziano1979}.

\section{Mass spectrum for $T>\tc$ in the case $N_f=2$}

As in the previous section, we start considering the range of temperatures
$\tro<T<\tuone$, with the parameters $\rho_\pi$ and $\rho_X$ given by
Eq. \eqref{Trange} (see also Table \ref{table2}).
We shall use for the field $U$ a more convenient variant of the linear
parametrization, while, for the field $X$, we shall use the
usual nonlinear parametrization given in Eq. \eqref{paramUXL3},
\begin{equation}\label{parUX}
U = \frac{1}{\sqrt2} [ (\sigma+i\eta)\mathbf{I}
+ (\vec{\delta}+i\vec{\pi})\cdot\vec{\tau} ] ,~~~~
X = \alpha e^{i\beta} ,
\end{equation}
where $\tau^a$ ($a=1,2,3$) are the three Pauli matrices [with the usual
normalization $\Tr(\tau^a\tau^b)=2\delta_{ab}$] and the fields
$\sigma$, $\eta$, $\vec\delta$, and $\vec\pi$ describe, precisely,
the $q\bar q$ mesonic excitations which are listed in Table \ref{table1}.

Inserting Eq. \eqref{parUX} and $M = {\rm diag}(m_u,m_d)$
into the expressions \eqref{V} and \eqref{Vtilde}, we find the following
expression for the potential with and without the anomalous term
(with $\omega_1=0$),
\begin{equation}\label{Vanom-bis}
\tilde{V} = V-\frac{1}{8}A[\log X-\log X^{\dagger}]^{2}
= V+\frac{1}{2}A\beta^2 ,~~~~~~~~~~
\end{equation}
\begin{equation}\label{VL2}
\nonumber
\begin{split}
V=&~\frac{1}{4}\lpq \Tr[(UU^\dagger)(UU^\dagger)]
+\frac{1}{4}\lambda_{\pi}^{\prime 2}[\Tr(UU^{\dagger})]^{2} \\
&+\frac{1}{4}\lambda_{\pi}^{2}B_{\pi}^{2}[\sigma^{2}+\eta^{2}
+\vec{\delta}^{2}+\vec{\pi}^{2}]
+\frac{1}{4}\lxq\left(\alpha^2-\frac{1}{2}F_X^2\right)^{2} \\
&-\frac{B_{m}}{2}[(m_{u}+m_{d})\sigma+(m_{u}-m_{d})\delta_{3}] \\
&-\frac{c_{1}}{2\sqrt2}
[\alpha \cos\beta(\sigma^{2}-\eta^{2}-\vec{\delta}^{2}+\vec{\pi}^{2}) \\
&+2\alpha \sin\beta(\sigma\eta-\vec{\delta}\cdot\vec{\pi})]
+\frac{1}{8}\lpq B_\pi^4 .
\end{split}
\end{equation}
When studying the equations for a {\it stationary point} of the potential,
one immediately finds that $\ov\eta = \ov\pi_a = \ov\beta = 0$ ($P$-invariance
requires that $\ov{U}=\ov{U}^\dagger$ and $\ov{X}=\ov{X}^\dagger$),
and also $\ov\delta_1 = \ov\delta_2 = 0$, while for the other values
$\ov\alpha$, $\ov\sigma$ and $\ov\delta \equiv \ov\delta_{3}$
one finds the following solution (at the first nontrivial order in the
quark masses):
\begin{eqnarray}\label{solution1}
\smin &=& \frac{B_m}{\lpq\bpq-c_1F_X} (m_u+m_d)+\dots, \nonumber \\
\dmin &=& \frac{B_m}{\lpq\bpq+c_1F_X} (m_u-m_d)+\dots, \nonumber \\
\overline{\alpha} &=& \frac{F_{X}}{\sqrt2} + {\cal O}(m^2),
\end{eqnarray}
which, in the chiral limit $m_u=m_d=0$, reduces to
\begin{equation}\label{solution1chiral}
\ov U = 0,~~~~ \ov X = \ov\alpha = \frac{F_X}{\rad2} ,
\end{equation}
signalling that the $SU(2)_L\otimes SU(2)_R$ chiral symmetry is restored,
while the $U(1)$ axial symmetry is broken by the $U(1)$ axial condensate
$\ov X$.

Studying the matrix of the second derivatives ({\it Hessian}) of the
potential with respect to the fields at the stationary point,
one finds that (in the chiral limit $m_u = m_d = 0$) there are (as in the
case $N_f \geq 3$) two {\it exotic} $0^\pm$ singlet mesonic excitations,
described by the fields $h_X = \alpha - \ov\alpha$ and $S_X = \ov\alpha \beta$,
with squared masses
$M^2_{h_X}=\lxq\fxq$, $M^2_{S_X} = \frac{A}{\ov{X}^2} = \frac{2A}{\fxq}$,
and, moreover, two $q\bar q$ chiral multiplets appear in the mass spectrum
of the effective Lagrangian, namely,
\begin{eqnarray}\label{multiplets}
(\sigma,\vec\pi):&~
M_{\sigma}^{2} = M_{\pi}^{2} = \frac{1}{2}(\lpq\bpq - \sqrt{2} c_1 \ov{X}),
\nonumber \\
(\eta,\vec\delta):&~
M_{\eta}^{2} = M_{\delta}^{2} = \frac{1}{2}(\lpq\bpq + \sqrt{2} c_1 \ov{X}),
\end{eqnarray}
signalling the restoration of the $SU(2)_L \otimes SU(2)_R$ chiral
symmetry.\footnote{From the results \eqref{multiplets} we see that the
stationary point \eqref{solution1chiral} is a minimum of the potential,
provided that $\lpq\bpq > c_1 F_X$; otherwise, the Hessian evaluated
at the stationary point would not be positive definite.
Remembering that, for $\tro < T < \tuone$,
$\rho_\pi \equiv -\frac{1}{2} \bpq < 0$, the condition for the stationary
point \eqref{solution1chiral} to be a minimum can be written as
$\Gpi \equiv c_1 F_X + 2\lpq \rho_\pi = c_1 F_X - \lpq\bpq < 0$.
In other words, assuming $c_1 F_X > 0$ and approximately constant (as a
function of the temperature $T$) around $\tro$, we have that the stationary
point \eqref{solution1chiral} is a solution, i.e., a minimum of the potential,
not immediately above $\tro$, where the parameter $\rho_\pi$ vanishes
(see Table \ref{table2}) and $\Gpi$ is positive, but (assuming that $\lpq\bpq$
becomes large enough increasing $T$, starting from $\lpq\bpq=0$ at $T=\tro$)
only for temperatures that are sufficiently higher than $\tro$, so that
the condition $\Gpi<0$ is satisfied, i.e., only for $T > \tc > \tro$, where
$\tc$ is defined by the condition $\Gpi(T=\tc) = 0$,
and it is just what we can call the {\it chiral transition temperature}.}
Instead, the squared masses of the $q\bar q$ mesonic excitations belonging
to a same $U(1)$ chiral multiplet, such as $(\sigma,\eta)$ and
$(\vec\pi,\vec\delta)$, remain {\it split} by the quantity
\begin{equation}\label{Msplit-bis}
\Delta M_{U(1)}^2 \equiv M_{\eta}^2-M_{\sigma}^2
= M_{\delta}^{2}-M_{\pi}^{2} = \sqrt{2} c_1 \ov{X},
\end{equation}
proportional to the $U(1)$ axial condensate $\ov X = \frac{F_X}{\rad2}$.
This result is to be contrasted with the corresponding result obtained
in the previous section for the $N_f\geq 3$ case, see Eq. \eqref{massesL3},
in which {\it all} (scalar and pseudoscalar) $q\bar q$ mesonic excitations
(described by the field $U$) turned out to be degenerate, with squared masses
$M_U^{2}=\frac{1}{2}\lpq\bpq$.

\section{Comments on the results and conclusions}

The difference in the mass spectrum of the $q\bar q$ mesonic excitations
(described by the field $U$) for $T>T_c$ between the case $N_f=2$ and the
case $N_f\geq 3$ is due to the different role of the interaction term
$\mathcal{L}_{int} = c_{I}[\det U+\det U^{\dagger}] + \dots$, with
$c_{I}\equiv\frac{c_1\ov X}{2\rad2}$, in the two cases.
When $N_f=2$, this term is (at the lowest order) quadratic in the
fields $U$ so that it contributes to the squared mass matrix.
Instead, when $N_f\geq 3$, this term is (at the lowest order) an interaction
term of order $N_f$ in the fields $U$ (e.g., a {\it cubic} interaction term
for $N_f=3$) so that, in the chiral limit, when $\ov{U}=0$, it does not
affect the masses of the $q\bar q$ mesonic excitations.

\begin{figure}[htbp]
\begin{center}
\includegraphics[width=1.9cm,angle=270]{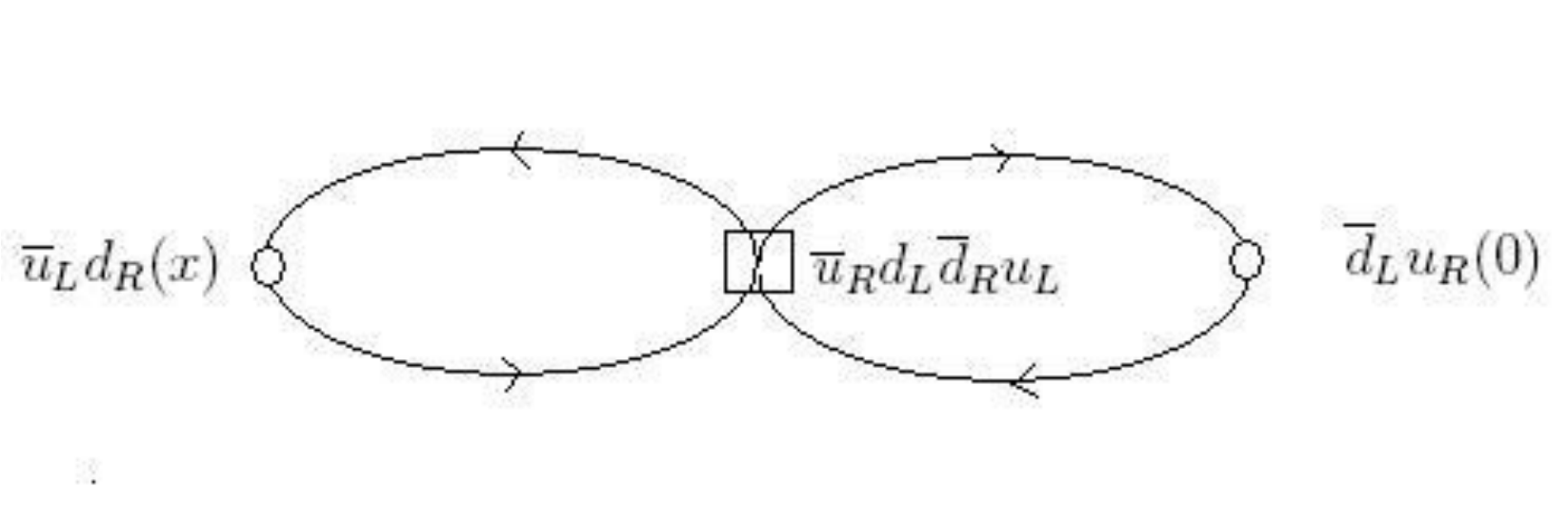}
\caption{Diagram with the contribution to ${\cal D}_{U(1)}$ from the
$2N_f$-quark effective vertex in the case $N_f=2$.}\label{fig1}
\end{center}
\end{figure}

\begin{figure}[htbp]
\begin{center}
\includegraphics[width=2.3cm,angle=270]{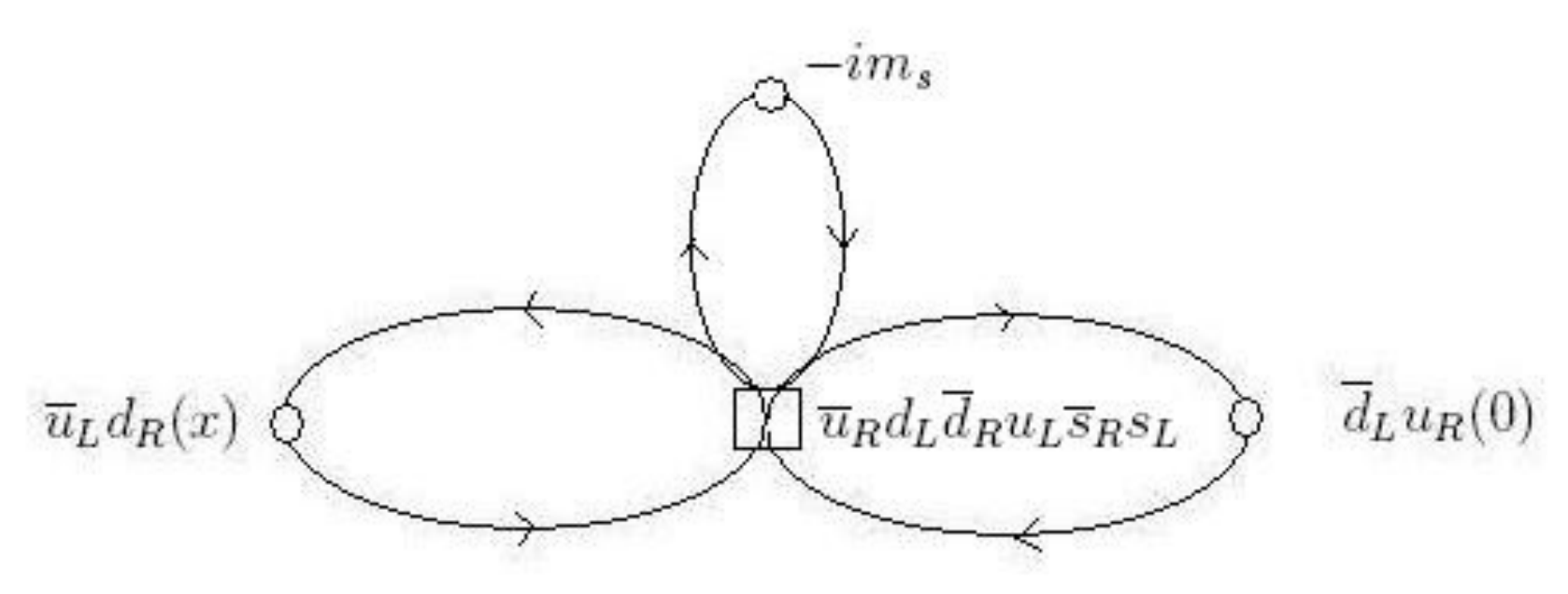}
\caption{Diagram with the contribution to ${\cal D}_{U(1)}$ from the
$2N_f$-quark effective vertex in the case $N_f=3$.}\label{fig2}
\end{center}
\end{figure}

Alternatively, we can also explain the difference by using a ``diagrammatic''
approach, i.e., by considering, for example, the diagrams that contribute
to the following quantity $\mathcal{D}_{U(1)}$, defined as the difference
between the correlators for the $\delta^+$ and $\pi^+$ channels:
\begin{equation}\label{DU1}
\begin{split}
{\cal D}&_{U(1)}(x) \equiv \langle T O_{\delta^+}(x) O_{\delta^+}^\dagger(0)
\rangle - \langle T O_{\pi^+}(x) O_{\pi^+}^\dagger(0) \rangle \\
&= 2 \left[ \langle T \bar{u}_R d_L(x) \ \bar{d}_R u_L(0) \rangle
+ \langle T \bar{u}_L d_R(x) \ \bar{d}_L u_R(0) \rangle \right] .
\end{split}
\end{equation}
What happens below and above $T_c$?
For $T<T_c$, in the chiral limit $m_1 = \dots ~m_{N_f} = 0$, the left-handed
and right-handed components of a given light quark flavor can be connected
through the $q\bar{q}$ chiral condensate, giving rise to a nonzero
contribution to the quantity ${\cal D}_{U(1)}(x)$ in Eq. \eqref{DU1}.
But for $T>T_c$, the $q\bar{q}$ chiral condensate is zero, and, therefore,
also the quantity ${\cal D}_{U(1)}(x)$
should be zero for $T>T_c$, {\it unless} there is a nonzero
$U(1)$ axial condensate $\ov{X}$; in that case, one should also consider
the diagram with the insertion of a $2N_f$-quark effective vertex
associated with the $U(1)$ axial condensate $\ov{X}$.
For $N_f=2$ (see Figure \ref{fig1}), all the left-handed and right-handed
components of the {\it up} and {\it down} quark fields in Eq. \eqref{DU1}
can be connected through the four-quark effective vertex,
giving rise to a nonzero contribution to the quantity ${\cal D}_{U(1)}(x)$.
Instead, for $N_f=3$ (see Figure \ref{fig2}), the six-quark effective vertex
also generates a couple of right-handed and left-handed {\it strange} quarks,
which, for $T>T_c$, can only be connected through the mass operator
$-m_s \ov{q}_s q_s$, so that (differently from the case $N_f=2$)
this contribution to the quantity ${\cal D}_{U(1)}(x)$
should vanish in the chiral limit; this implies that, for $N_f=3$ and $T>T_c$,
the $\vec\delta$ and $\vec\pi$ correlators are identical,
and, as a consequence, also $M_\delta = M_\pi$.
This argument can be easily generalized to include also the other meson
channels and to the case $N_f>3$.

Finally, let us see how our results for the mass spectrum compare with
the available lattice results.
Lattice results for the case $N_f=2$ (and for the case $N_f=2+1$, with
$m_{u,d} \to 0$ and $m_s \sim 100$ MeV) exist in the literature, even if
the situation is, at the moment, a bit controversial.
In fact, almost all lattice results
\cite{lat1997,lat1998,lat1999,lat2000,lat2000bis,lat2011,lat2012,lat2014}
(using {\it staggered fermions} or {\it domain-wall fermions} on
the lattice) indicate the {\it nonrestoration} of the $U(1)$ axial
symmetry above the chiral transition at $T_c$, in the form of a small (but
nonzero) splitting between the $\vec\delta$ and $\vec\pi$ correlators above
$T_c$, up to $\sim 1.2~T_c$.
In terms of our result \eqref{Msplit-bis}, we would interpret this by saying
that, for $T>\tc$, there is still a nonzero $U(1)$ axial condensate,
$\ov{X} > 0$, so that $c_{I}=\frac{c_1\ov X}{2\rad2} > 0$ and the
above-mentioned interaction term, containing the determinant of the mesonic
field $U$, is still effective for $T>\tc$.

However, other lattice results obtained in Ref. \cite{Cossu2013}
(using the so-called {\it overlap fermions} on the lattice; see also
Ref. \cite{Aoki2012})
do not show evidence of the above-mentioned splitting above $\tc$, so
indicating an {\it effective} restoration of the $U(1)$ axial symmetry
above $\tc$, at least, at the level of the $q\bar{q}$ mesonic mass spectrum.
In terms of our result \eqref{Msplit-bis}, we would interpret this by saying
that, for $T>\tc$, one has $c_1 \ov{X} = 0$, so that
$c_{I}=\frac{c_1\ov X}{2\rad2} = 0$ and the above-mentioned interaction term,
containing the determinant of the mesonic field $U$, is not present
for $T>\tc$.
For example, it could be that also the $U(1)$ axial condensate $\ov{X}$
(like the usual chiral condensate $\langle \bar{q} q \rangle$) vanishes
at $T=\tc$, i.e., using the notation introduced in Sec. 2
(see Table \ref{table2}), that $\tuone=\tc$. (Or, even more drastically, it
could be that there is simply {\it no} genuine $U(1)$ axial condensate \dots)

In conclusion, further work will be necessary, both from the analytical point
of view but especially from the numerical point of view (i.e., by lattice
calculations), in order to unveil the persistent mystery of the fate of the
$U(1)$ axial symmetry at finite temperature.

Also the question of the (possible) {\it exotic} pseudoscalar singlet field
$S_X \sim i[ {\det}(\bar{q}_{sL}q_{tR}) - {\det}(\bar{q}_{sR}q_{tL})]$
for $T>\tc$, with squared mass (in the chiral limit) given by
$M^2_{S_X}|_{M=0} = \frac{A}{\ov{X}^2} = \frac{2A}{\fxq}$,
should be further investigated, both theoretically and experimentally.
As we have already said, the excitation $S_X$ is nothing but the
{\it would-be} Goldstone particle coming from the breaking of the $U(1)$
axial symmetry, as required by the Witten--Veneziano mechanism
\cite{Witten1979,Veneziano1979}.
So, it is precisely what we should call the ``$\eta'$'' for $T>\tc$:
is there any chance to observe it?
Lattice results seem to indicate that $A(T)$ has a sharp decrease for $T>\tc$
and it vanishes at $\sim 1.2~T_c$ \cite{VP-report}.
(And, maybe, $A(T>\tc) \to 0$ for $N_c \to \infty$,
as it was suggested in Ref. \cite{KPT1998}.)
Could this explain the ``$\eta'$'' mass decrease, which, according
to Ref. \cite{Csorgo-et-al.2010}, has been observed inside the fireball
in heavy-ion collisions?







\end{document}